\documentclass[prd,aps,nofootinbib,preprint,showpacs,showkeys,preprintnumbers,amssymb,superscriptaddress]{revtex4}
\usepackage[english]{babel}
\usepackage{amsmath}
\usepackage[dvips]{graphicx}
\usepackage{ifthen,array}
\usepackage[sort&compress]{natbib}
\usepackage{amsfonts}
\usepackage{bbm}
\usepackage{dcolumn}
\usepackage{bm}

\DeclareMathAlphabet{\mathsc}{OT1}{cmr}{m}{sc}

\allowdisplaybreaks

\def\10{$SO(10)$}
\def\21{SU(2) $\otimes$ U(1) }

\def\422{$SU(4) \otimes SU(2) \otimes SU(2)$}
\def\321{SU(3) $\otimes$ SU(2) $\otimes$ U(1)}

\def\lsim{\raise0.3ex\hbox{$\;<$\kern-0.75em\raise-1.1ex\hbox{$\sim\;$}}}
\def\gsim{\raise0.3ex\hbox{$\;>$\kern-0.75em\raise-1.1ex\hbox{$\sim\;$}}}

\newcommand{\AddrAHEP}{%
  AHEP Group, Institut de F\'{\i}sica Corpuscular --
  C.S.I.C./Universitat de Val{\`e}ncia \\
  Edificio Institutos de Paterna, Apartado: 22085, E--46071 Valencia, Spain}

\newcommand{\AddrLisb}{%
 Departamento de F\'\i sica and CFTP, Instituto Superior T\'ecnico\\
          Avenida Rovisco Pais 1, 1049-001 Lisboa, Portugal }
\begin{document}

\preprint{IFIC/07-28}
\preprint{SU 4252-831}

\title{Exact relativistic beta decay endpoint spectrum} 
\author{S.~S.~Masood} \email{masood@uhcl.edu}
\affiliation{Department of Physics, 2700 Bay Area Blvd., University of
  Houston Clear Lake, Houston, TX 77058, USA} 
\author{S.~Nasri} \email{snasri@ufl.edu} \affiliation{Institute for Fundamental
Theory, University of Florida, Gainesville, FL, USA} 
\author{J.~Schechter}
\email{schechte@phy.syr.edu} \affiliation{Department of
  Physics, Syracuse University, Syracuse, NY 13244-1130, USA}
\author{M.~ A.~T\'ortola}\email{mariam@cftp.ist.utl.pt}
\affiliation{\AddrLisb} 
\author{J.~W.~F.~Valle}\email{valle@ific.uv.es} \affiliation{\AddrAHEP} 
\author{C.~Weinheimer} \email{weinheimer@uni-muenster.de} \affiliation{Institut
  f\"ur Kernphysik, Westf\"alische Wilhelms-Universit\"at M\"unster\\
  Wilhelm-Klemm-Str. 9; D-48149 M\"unster, Germany}

\date{\today}

\begin{abstract}
The exact relativistic form for the beta decay endpoint spectrum is
derived and presented in a simple factorized form.  We show that our 
exact formula can be well approximated to yield the endpoint
form used in the fit method of the KATRIN collaboration.  We also
discuss the three neutrino case and how information from neutrino
oscillation experiments may be useful in analyzing future beta decay
endpoint experiments.

\end{abstract}
\maketitle

\section{Introduction}

The discovery of neutrino
oscillations~\cite{fukuda:2002pe,ahmad:2002jz,araki:2004mb,Kajita:2004ga,ahn:2002up}
probes the neutrino squared mass differences and mixing
angles~\cite{Maltoni:2004ei}, but leaves open the issue of what is the
absolute scale of neutrino mass. The latter has important cosmological
implications in the cosmic microwave background and large scale
structure in the Universe, as already indicated by the sensitivities
reached, for example, by the recent WMAP-3~\cite{Spergel:2006hy}, the
2dF Galaxy Redshift Survey~\cite{colless2001grs} and Sloan Digital Sky
Survey results~\cite{Tegmark:2003ud}.
One expects even better sensitivities in the next generation of
cosmological observations~\cite{Lesgourgues:2006nd,Hannestad:2006zg}.
Interesting as these may be, there are essentially only two ways to
get insight into the absolute scale of neutrino mass in the
laboratory: searches for neutrinoless double beta
decay~\cite{elliott:2002xe} and investigations of the beta spectra
near their endpoints
\cite{osipowicz:2001sq,Lobashev:2003kt,Kraus:2004zw,Drexlin:2005zt,weinheimer07,Angrik:2005ep}.
For the latter direct search for the neutrino mass a very low beta
endpoint is crucial: tritium was used in the most sensitive
spectrometer experiments \cite{Kraus:2004zw,Lobashev:2003kt} and
rhenium in the up-coming cryobolometer experiments
\cite{Monfardini:2005dk}.
 
Currently a next generation tritium beta-decay experiment is being
prepared, scaling up the size and precision of previous experiments by
an order of magnitude, and increasing the intensity of the tritium
beta source: the KArlruhe TRItium Neutrino experiment
KATRIN~\cite{osipowicz:2001sq,Drexlin:2005zt,weinheimer07,Angrik:2005ep}.
Such an improved sensitivity experiment will probe neutrino masses 
ten times smaller than the current limits and therefore play a crucial
role in probing for direct effects of neutrino masses.

Prompted by the prospects that high sensitivities can be achieved in
the next generation of high precision neutrino mass searches from
tritium beta decay experiments~\cite{Drexlin:2005zt,weinheimer07} we
reexamine the accuracy of the kinematical formulae used in the
determination of neutrino masses from the shape of the endpoint
spectrum. We also discuss the interplay of neutrino oscillation data
and the expectations for the beta decay endpoint counting rates for
the different types of neutrino mass spectra.

\section{Relativistic beta decay kinematics}

In what follows we label the relativistic momenta and energies
involved in tritium beta decay according to
\begin{equation}
  \label{eq:A}
  ^3 H (\bm{0}, M) \to   ^3 {He}^+ (\bm{p^\prime}, E^\prime) +
e^- (\bm{p_e}, E_e) + \bar{\nu_e} (\bm{p_\nu}, E_\nu). 
\end{equation}
The masses of $^3 {He}^+$, $e^-$ and $\bar{\nu_e}$ are denoted by
$M^\prime$, $m_e$ and $m_\nu$ respectively. In order to see the
convenience of an exact relativistic description we mention, as
recently noted in Ref.~\cite{Masood:2005aj} that the well known
relativistic formula for the maximum electron energy
\begin{equation}
  \label{eq:B}
E_e^{max} = \frac{1}{2M}\left[M^2 + m_e^2 - (m_\nu + M^\prime)^2\right],
\end{equation}
gives a value about 3.4 eV lower than the approximation
$M-M^\prime-m_\nu$ often used. This suggests the desirability of carrying
out the full phase space integration using relativistic kinematics.

Start from the standard formula for the decay width at rest,
\begin{equation}
  \label{eq:C}
\Gamma = \frac{1}{2^9\pi^5 M} \int \frac{d^3p_e d^3p_\nu 
d^3p^{\prime}}{E_e E_\nu E'}
 |{\cal{M}}|^2 \delta^4(p_{\mathrm{initial}}-p^\prime-p_e-p_\nu),
\end{equation}
where $|{\cal M}|^2$ denotes the spin-summed, Lorentz invariant
``squared'' amplitude. To explore the constraint of Lorentz invariance
one might a priori consider expanding $|{\cal M}|^2$ in terms of
invariants constructed from the four-momenta. For example, up to two
powers of momenta, the most general form is,
\begin{equation}
  \label{eq:D}
 |{\cal{M}}|^2 = A - B p_e\cdot p_\nu  - C p'\cdot p_{\mathrm{initial}} + ...,
\end{equation}
where $A, B$ and $C$ are constants. Now it is easy to perform 
some initial
integrations. As usual $\int d^3p^\prime$ is first done with the
momentum delta-function. Then the angle between ${\bf p}_e$ and 
${\bf p}_\nu$ is
eliminated using the energy delta function. Three more
angular integrals are trivial.
As the result 
 one may replace in the ${}^3H$ rest frame
\begin{equation} 
|{\cal{M}}|^2  \to A + B( E_e E_\nu -{\bf p_e}\cdot{\bf p_\nu})  
+CM(M-E_e-E_\nu),
\label{eqn:E}
\end{equation}  
where
\begin{equation}
{\bf p_e}\cdot{\bf 
p_\nu} \equiv \frac{1}{2} \left[ M^2-M'^2+m_e^2+m_\nu^2-2ME_e+2E_\nu(E_e-M) \right].
\label{enuangle}
\end{equation}
Eq.~(\ref{eqn:E}) can now be inserted in the resulting usual
formula~\cite{Eidelman:2004wy}
\begin{equation}
  \label{eq:F}
 \Gamma = \frac{1}{2^6\pi^3 M}\int dE_\nu dE_e |{\cal{M}}|^2 .
\end{equation}
Next we find $d\Gamma/dE_e$ by integrating over $dE_\nu$ for each
$E_e$. The limits of integration $E_\nu^{min} (E_e)$ and $E_\nu^{max}
(E_e)$ can be read from \cite{Eidelman:2004wy}. The most tedious part
of the present calculation is finding the factorizations:
\begin{equation}
  \label{eq:G}
 E_\nu^{max} - E_\nu^{min} = \frac{2Mp_e}{(m_{12})^2}(E_e^{max} - E_e)^{1/2}
\left[E_e^{max} - E_e + \frac{2m_\nu M'}{M}\right]^{1/2},
\end{equation}
\begin{equation}
  \label{eq:H}
 E_\nu^{max} + E_\nu^{min} = \frac{2M}{(m_{12})^2}(M - E_e)
\left[E_e^{max} - E_e + \frac{m_\nu}{M}(M'+m_\nu) \right],
\end{equation}
wherein:
\begin{equation}
  \label{eq:I}
 (m_{12})^2 = M^2 - 2ME_e + m_e^2.
\end{equation}
The importance of the factorization is that it makes the behavior at
the endpoint $E_e=E_e^{max}$ transparent. Then we have the exact 
relativistic result,
\begin{eqnarray}
  && \frac{d\Gamma}{dE_e} =
  \frac{1}{(2\pi)^3}\frac{p_e}{4(m_{12})^2} \sqrt{y\left(y + \frac{2
      m_\nu M'}{M}\right)}[A + CM(M-E_e) +   
\nonumber \\ 
  &&  +   BM\frac{ME_e-m_e^2}{(m_{12})^2}\left(y +
 \frac{m_\nu}{M}(M' + m_\nu)\right) 
  -C\frac{M^2}{(m_{12})^2}(M-E_e)\left(y +
 \frac{m_\nu}{M}(M' + m_\nu)\right)],
  \label{eq:J}
\end{eqnarray}
where $y=E_e^{max}- E_e$.

As it stands, this formula is based only on the kinematical assumption
in Eq.~(\ref{eq:D}). It obviously vanishes at the endpoint $y=0$ as
$\sqrt y$.  Note that all other terms are finite at $y=0$. The overall
factor $\sqrt{y(y+2m_\nu M^\prime/M)}$ gives the behavior of $\frac{d
  \Gamma}{dE_e}$ extremely close to $y=0$ for any choice of A, B and
C, but departs from $\frac{d \Gamma}{dE_e}$ away from the endpoint.

Dynamics is traditionally put into the picture~\cite{bjorken:xxxx} by
examining the spin sum for a 4-fermion interaction wherein the nuclear
matrix element is assumed constant. This is presented as a 
non- Lorentz invariant term,
\begin{equation}
|{\cal{M}}|^2=BE_eE_\nu.
\label{tradition}
\end{equation}
 We will see that this is excellently approximated in our fully
relativistic model by,
\begin{equation}
  \label{eq:L}
A=C=0, \:  \: \: \:  B\neq 0.
\end{equation}
  A more accurate treatment of the underlying
interaction might give rise to small admixtures of non-zero A
and C as well as other unwritten coefficients in Eq.~(\ref{eq:D})
above.

The form for the spectrum shape near the endpoint that results from
putting $A=C=0$ in Eq.~(\ref{eq:J}) is
\begin{equation}
  \label{eq:M}
\frac{d\Gamma}{dE_e} = \frac{p_eMB}{(2\pi)^3 4(m_{12})^4}
(ME_e-m_e^2)\sqrt{y\left(y + \frac{2 m_\nu M'}{M}\right)} \left[ y + 
\frac{m_\nu}{M}(M' + m_\nu)\right].
\end{equation}
Note that if we had employed the non-relativistic form given
in Eq.~(\ref{tradition}) the net result would be a replacement of
an overall factor in Eq.~(\ref{eq:M}) according to,
\begin{equation}
(ME_e-m_e^2)\rightarrow (ME_e-E_e^2).
\label{replacement}
\end{equation}
The difference of these two factors yields the contribution of
the ${\bf p_e}\cdot{\bf p_\nu}$ term. It is  
really
negligible near the endpoint region since it is proportional to $p_e^2$
and is suppressed like $p_e^2/(ME_e)$ compared to unity.
We have checked that the result of our calculation with just the 
$E_eE_\nu$ term  agrees with the calculation
of Ref.~\cite{Wu:1983sz}, though their result looks much more
complicated, as they did not present it in the simpler
factorized form given here.

Note that only the two rightmost factors vary appreciably near the
endpoint of Eq.~(\ref{eq:M}). If we further approximate $M^\prime/M
\to 1$ and $\frac{M^\prime + m_\nu}{M} \to 1$ the endpoint shape is
well described by
\begin{equation}
  \label{eq:N}
\frac{d\Gamma}{dE_e} \quad \alpha \quad (y + m_\nu) \sqrt{y(y+2m_\nu)}.
\end{equation}
Now we compare with the formula used in the experimental
analysis~\cite{Kraus:2004zw}
\begin{equation}
  \label{eq:O}
\frac{d\Gamma}{dE_e} \quad  \alpha \quad (E_0 - V_i - E)\sqrt{(E_0-V_i-E)^2
 - m_\nu^2}.
\end{equation}
This agrees with the above approximation in Eq.~(\ref{eq:N}) if
one identifies
\begin{equation}
  \label{eq:P}
(E_0 - V_i - E) = y + m_\nu.
\end{equation}
Note that $E$ is the non-relativistic energy given by $E=E_e-m_e$.
Furthermore, $E_0-V_i$ is identified with our $(M-M^\prime -m_e-
\delta E_e^{max})$. $\delta E_e^{max}$ is defined by,
\begin{equation}
  \label{eq:Q}
E_e^{max} = M-M^{\prime}- m_\nu -\delta E_e^{max},
\end{equation}                                                            
and was shown in \cite{Masood:2005aj} to be independent of $m_\nu$ to
a good approximation. Thus we see that the exact relativistic
endpoint structure obtained here may be well approximated by the form
used in the experimental analysis.

\begin{figure}[t]
  \centering
  \includegraphics[width=12cm,height=6cm]{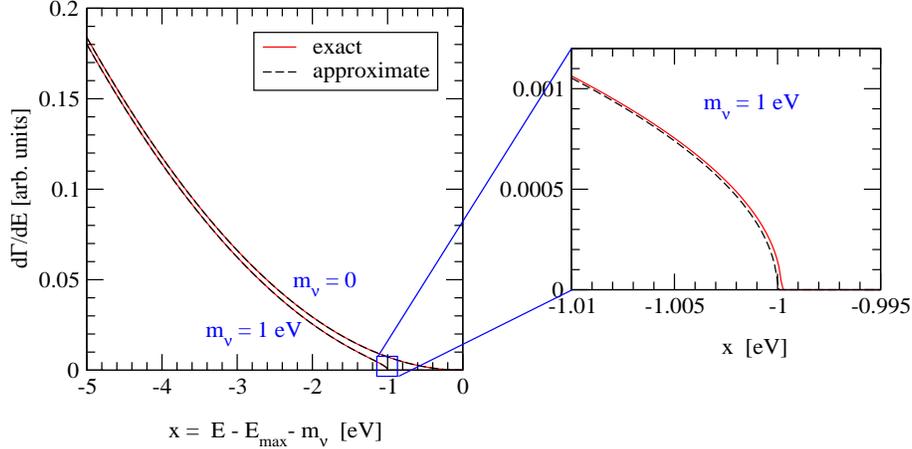}
  \caption{Comparison of approximate and exact formula}
  \label{fig:comp}
\end{figure}

    Often, authors express results in terms of a variable, $x$, which from 
our discussion may be seen to be the same as,
\begin{equation}
x=-y-m_{\nu}=E_e-E_e^{max}-m_{\nu}.
\label{definex}
\end{equation}
In Fig.~\ref{fig:comp}, $d\Gamma/dE_e$ as computed from the exact
formula, Eq.~(\ref{eq:M}) is compared with its approximate analog as a
function of $x$.  As can be seen, the differences between the
approximate and exact formulae are tiny.

It may be worthwhile to remark that the exact relativistic kinematical
expression in Eq.~(\ref{eq:M}) is no more complicated than the
approximation one ordinarily uses.

\section{Numerical Simulations}
\label{sec:numer-simul}

We have done Monte Carlo simulations by creating random data sets
following the exact relativistic kinematical expression in
Eq.~(\ref{eq:M}) for neutrino masses of $m_\nu = 0$~eV and $m_\nu =
1$~eV and fitting them with the standard formula Eq.~(\ref{eq:O}). The
simulations were performed for a KATRIN-like experiment
\cite{Angrik:2005ep} considering:
\begin{itemize}

\item The ro-vibrational states of the $T_2 \rightarrow (T^3He)^+$
  decay \cite{PhysRevLett.84.242}
\item A signal rate from a KATRIN-like molecular gaseous windowless
  tritium source with a column density of $5 \cdot 10^{17}$
  molecules/cm$^2$ over an active area of 53~cm$^2$ and an accepted
  solid angle of $\Delta \Omega/4\pi = 0.18$
\item An expected background rate of 0.01~s$^{-1}$.   
\item A response function of a KATRIN-like experiment considering the 
  energy losses within the tritium source
  and the main spectrometer transmission function with a total width of 0.93~eV.  
\item 3 years of total data taking covering an energy range of the
  25~eV below and 5~eV above the tritium endpoint following an
  optimized measurement point distribution \cite{Angrik:2005ep}.
\end{itemize}

\begin{figure}[t]
  \centering
  \includegraphics[width=14cm]{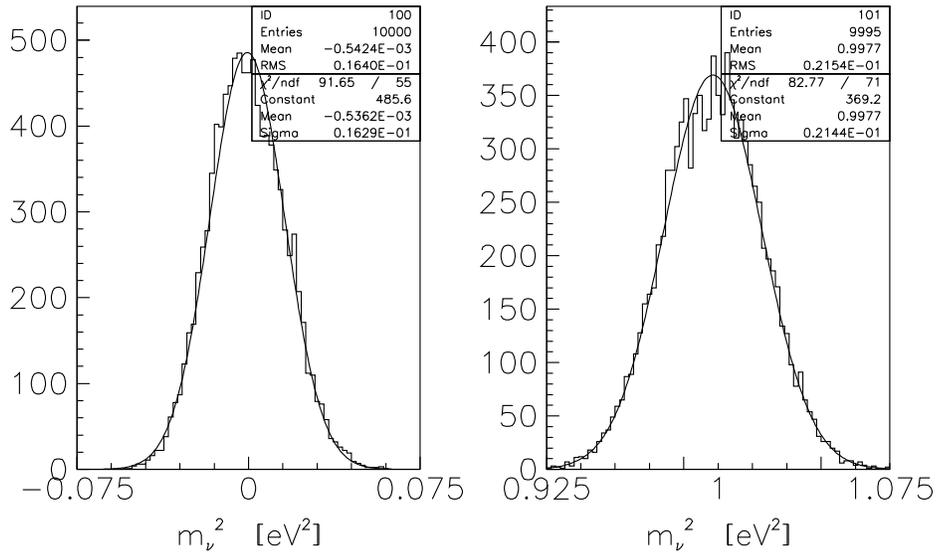}
  \caption{Results on the observable $m^2_\nu$ from the fitting of
    10000 sets of Monte Carlo data randomized according to the exact
    relativistic formula Eq.~(\ref{eq:M}) and to a fitting routine
    based on the standard formula Eq.~(\ref{eq:O}), for neutrino masses
    of 0~eV (left) and 1~eV (right). The rms values of the
    Gaussian-like distributions correspond to the expected statistical
    uncertainty $\Delta m^2_{\nu, stat}$ for a KATRIN-like
    experiment.}
  \label{fig:montecarlo_fits}
\end{figure}

Fig. \ref{fig:montecarlo_fits} shows the results for the observable
$m^2_\nu$ obtained from the fitting of 10000 sets of Monte Carlo data
randomized according to the exact relativistic formula Eq.~(\ref{eq:M})
and to a fitting routine using the standard formula Eq.~(\ref{eq:O}),
assuming neutrino masses of 0~eV (left) and 1~eV (right).
The rms values of the Gaussian-like distributions correspond to the
expected statistical uncertainty $\Delta m^2_{\nu, stat}$ for a
KATRIN-like experiment.
Clearly the mean value of the fit results for the neutrino mass
squared $m_\nu^2$ does not show any significant deviation from the
starting assumption of $m_\nu = 0$~eV or $m_\nu = 1$~eV, respectively.
This establishes that the exact relativistic formula Eq.~(\ref{eq:M})
can be well approximated by the standard equation (\ref{eq:O}) for the
precision needed for the next generation tritium experiment KATRIN.
This is probably due to the fact that KATRIN is investigating the last
25~eV below of the beta spectrum below its endpoint only, where the
recoil corrections are nearly independent on the electron energy.

\section{Three neutrino case}

Of course, the most interesting application is to the case of three
neutrinos with different masses, $m_1$, $m_2$ and $m_3$. Then there
will be a different endpoint energy, $E_i^{max}$ corresponding to each
one. The effective endpoint factor in the good approximation of
Eq.~(\ref{eq:N}) is the weighted sum,
\begin{equation}
\label{F}
F_{eff}(E_e)=\sum_{i=1}^3|K_{1i}|^2(y_i+m_i)[y_i(y_i+2m_i)]^{1/2}
\theta(y_i),  
\end{equation}
where $y_i(E_e)=E_i^{max}-E_e$ and the $K_{1i}$ are the elements of
the 3x3 lepton mixing matrix~\cite{schechter:1980gr,Yao:2006px}.  We
note that the further good approximation that the quantity $\delta
E_e^{max}$ is independent of the neutrino mass, gives the useful
relation
\begin{equation}
\label{useful}
y_i-y_j=m_j-m_i.
\end{equation}
Now let the unindexed quantity $y$ stand for the $y_i$ with the
smallest of the neutrino masses.  Using Eq.~(\ref{useful}) allows us to
write the explicit formula for the case (denoted ``normal hierarchy'')
where $m_1$ is the lightest of the three neutrino masses as:
\begin{eqnarray}
F_{NH}(y)&=& |K_{11}|^2(y+m_1)[y(y+2m_1)]^{1/2} \\ \nonumber
      &+& 
|K_{12}|^2(y+m_1)[(y+m_1-m_2)(y+m_1+m_2)]^{1/2}\theta(y+m_1-m_2)\\
\nonumber
      &+& 
|K_{13}|^2(y+m_1)[(y+m_1-m_3)(y+m_1+m_3)]^{1/2}\theta(y+m_1-m_3).
\label{FI}
\end{eqnarray}
In the other case of interest (denoted ``inverse hierarchy'') we have:
\begin{eqnarray}
F_{IH}(y)&=& |K_{13}|^2(y+m_3)[y(y+2m_3)]^{1/2} \\ \nonumber
      &+&
|K_{11}|^2(y+m_3)[(y+m_3-m_1)(y+m_3+m_1)]^{1/2}\theta(y+m_3-m_1)\\
\nonumber
      &+&
|K_{12}|^2(y+m_3)[(y+m_3-m_2)(y+m_3+m_2)]^{1/2}\theta(y+m_3-m_2).
\label{FII}
\end{eqnarray}                                              
where $m_3$ is the lightest of the three neutrino masses. From these
equations we may easily find the counting rate in the energy range
from the appropriate endpoint up to $y_{max}$ as proportional to the
integral
\begin{equation}
n_{NH}(y_{max}) ={\int}_0^{y_{max}}dyF_{NH}(y),
\label{nI}
\end{equation}
or, for the ``inverse hierarchy'' case, as proportional to,
\begin{equation}
n_{IH}(y_{max}) ={\int}_0^{y_{max}}dyF_{IH}(y).
\label{nII}
\end{equation}                 

We note that, as stressed in ref.  \cite{Masood:2005aj}, information
on neutrino masses and mixings obtained from neutrino oscillation
experiments is actually sufficient in principle to predict
$n(y_{max})$ as a function of a single parameter (up to a twofold
ambiguity). Thus, in principle, suitably comparing the predicted
values of $n(y_{max})$ with results from a future endpoint experiment
may end up determining three neutrino masses.

To see how this might work out we make an initial estimate using the
best fit values \cite{Maltoni:2004ei} of neutrino squared mass
differences,
\begin{eqnarray} 
\label{AB}
A&\equiv& m_2^2-m_1^2=7.9\times10^{-5}eV^2, \\ \nonumber
B&\equiv&|m_3^2-m_2^2|=2.6\times10^{-3}eV^2,
\end{eqnarray}
and the weighting coefficients,
\begin{eqnarray}
|K_{11}|^2&=&0.67, \\ \nonumber
|K_{12}|^2&=&0.29, \\ \nonumber
|K_{13}|^2&=&0.04.
\label{Kij}
\end{eqnarray}
Currently $|K_{13}|^2$ is consistent with zero and is only bounded.
For definiteness we have taken a value close to the present upper
bound. However, we have checked that the effect of putting it to zero
is very small.  Now, from the two known differences in Eq.~(\ref{AB})
we can for each choice of $m_3$ (considered as our free parameter)
find the masses $m_1$ and $m_2$, subject to the ambiguity as to
whether $m_3$ is the largest (NH) or the smallest (IH) of the three
neutrino masses.  Of course we hope that future long baseline neutrino
oscillation
experiments~\cite{albright:2000xi,apollonio:2002en,Huber:2002mx,Alsharoa:2002wu}
might eventually determine whether nature prefers the NH or the IH
scenario.

\begin{figure}[t]
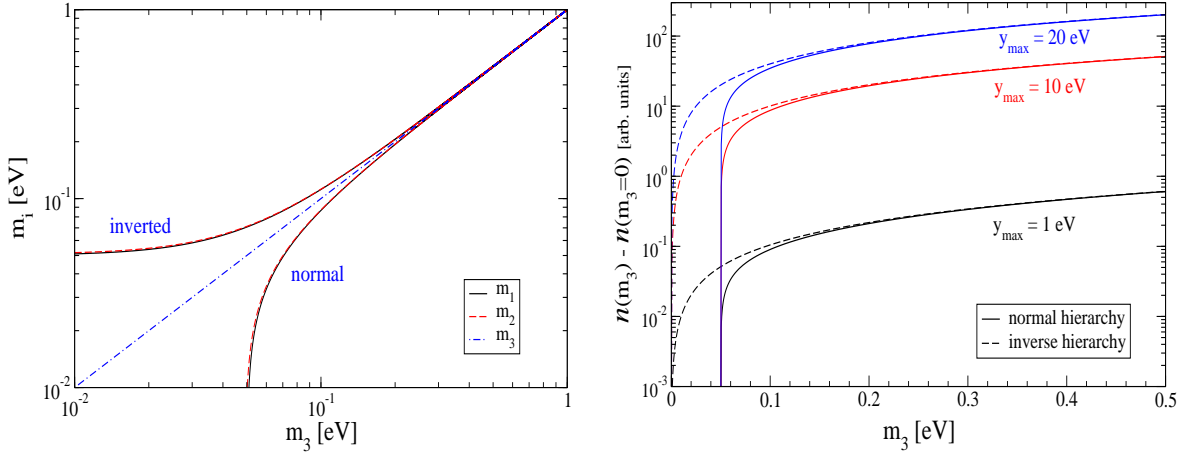

  \centering
   \includegraphics[width=7.5cm,height=6cm]{numass-m3-log.eps}
\hspace{3mm}
 \includegraphics[width=7.5cm,height=6cm]{count-m3-log.eps}
 \caption[]{The left panel shows typical solutions for $(m_1, m_2)$ as
   a function of $m_3$ for the NH case (solid curves) and the IH case
   (dashed curves); the middle dot-dashed is given for orientation.
   The right panel give the predictions for the quantities,
   $n(y_{max})$, proportional to the event counting rate which
   includes emitted electrons within, respectively 1 eV, 10 eV and 20
   eV from the appropriate endpoint. 
\label{fig:typical}}
\end{figure}

Fig.~\ref{fig:typical} shows typical solutions for the mass set
($m_1,m_2$) in terms of the free parameter $m_3$. Very large values of
$m_3$ would fall within the sensitivity of upcoming cosmological tests
\cite{Lesgourgues:2006nd,Hannestad:2006zg}.
In the right panel of Fig.~\ref{fig:typical} we display the predicted
values of $n(y_{max})$ for each possible mass scenario and the choices
of (1,10,20) eV for $y_{max}$.
These quantities are proportional to the electron counting rate in the
energy interval from the endpoint (for each mass scenario) to
$y_{max}$ eV below the endpoint. The different values of $y_{max}$
reflect, of course, different experimental sensitivities.  The main
point is that, for sufficiently large $m_3$ values, the counting rate
is seen to distinguish the different possible neutrino mass sets from 
each other.
We hope that the present method of relating observed neutrino
oscillation parameters to predictions for the beta decay endpoint
counting rates may play a useful role in the forthcoming experiments.

 \section{Summary and discussion}

 We have derived the exact relativistic form for the beta decay
 endpoint spectrum and presented it in a very simple and useful
 factorized form. 
 We showed that our exact formula can be well approximated to yield 
 the endpoint form used in the fit method of the KATRIN collaboration. 
 This was explicitly established through a detailed numerical simulation.
 We have also discussed the three neutrino case and shown how
 information from neutrino oscillation experiments may be useful in
 analyzing future beta decay endpoint
 experiments. 

\section*{Acknowledgements}

We are grateful to R. Shrock for helpful discussions.  S.~S.~M.  would
like to thank the Physics Department at Syracuse University for their
hospitality. J.~S. would like to express his appreciation for the
hospitality received from the AHEP Group of IFIC during the summer of
2005. This work was supported by Spanish grants FPA2005-01269 and by
the EC RTN network MRTN-CT-2004-503369.  The work of J.S. is supported
in part by the U.S. DOE under Contract no.  DE-FG-02-85ER 40231. S.N.
was supported by the DOE Grant DE-FG02-97ER41029.
                                                                    

\end{document}